# Composing Ensembles of Instrument-Model Pairs for Optimizing Profitability in Algorithmic Trading

Sahand Hassanizorgabad, KOC University, School of Sciences and Engineering, Turkey

**Abstract**

Financial markets are nonlinear with complexity, where different types of assets are traded between buyers and sellers, each having a view to maximize their Return on Investment (ROI). Forecasting market trends is a challenging task since various factors like stock-specific news, company profiles, public sentiments, and global economic conditions influence them. This paper describes a daily price directional predictive system of financial instruments, addressing the difficulty of predicting short-term price movements. This paper will introduce the development of a novel trading system methodology by proposing a two-layer Composing Ensembles architecture, optimized through grid search, to predict whether the price will rise or fall the next day. This strategy was back-tested on a wide range of financial instruments and time frames, demonstrating an improvement of 20% over the benchmark, representing a standard investment strategy.

---

# 1 INTRODUCTION

## 1.1 Background

A financial market is a nonlinear complex system in which different kinds of traders and investors trade various assets to maximize returns on investment [1]. Forecasting financial markets is continuouss and always challenging since the markets keep on dynamically and volatily changing due to many factors that include news about the company, macroeconomic events, public mood, and international economic conditions. These are interdependent variables, creating an environment in which forecasted trends are very difficult to achieve [2].

The years have thrown up, so to say, several forecasting methods from traditional statistical models to more modern techniques involving machine learning. However, the attainment of consistent, reliable prediction has remained elusive given the inherent stochasticity of financial data. More relevant to traders, investors, and financial institutions than the accurate prediction of the increase or decrease in the price of an asset on a day-to-day basis, even minor improvements in the direction of the prediction may yield substantial financial benefits.

This development represents one of the more important growths in finance: algorithmic trading [3]. The automation of the decision to trade comes via the power of computer algorithms according to predefined parameters. Operating near and beyond human capabilities in speed and frequency, they leverage historical and real-time data to predict near-term price movements.

Algorithmic trading has become an integral part of financial markets today, leveraging the speed and precision of automated systems to make trades based on a predetermined set of criteria. This form of trading revolutionized the way trading occurs, now allowing complicated strategies in real time that would have been impossible to be manually undertaken by humans. Algorithmic trading systems use an immense amount of historical and real-time data to forecast near-term price movements, hence enabling the trader to gain from market inefficiencies at speeds not even conceivable for human traders. The need for fast and exact trading decisions has made algorithmic trading a very crucial element of today's financial markets.

## 1.2 Research Problem

Despite all the progress in the development of algorithmic trading, the stumbling block remains the challenge of achieving high accuracy and profitability on a continuous basis. While most models in existence suffer from being focused on one instrument or some grouping of instruments, which makes them inflexible and adaptable only under different market conditions. Their achieved accuracy, though respectable in some cases, turns out to be disappointing with regard to the ROI. These models, in general, are fitted for specific scenarios or market phases and work well, but when it comes to long-term usability, they lack flexibility in conditions, hindering their performance over an extended period.

The major weakness of conventional algorithmic trading models is that they cannot adapt dynamically to a multitude of instruments and market circumstances. The majority of the papers are single-model or single-instrument approaches; hence, most of them lack generality. These systems generally work well on a small set of assets and fail when applied to other instruments or market phases, such as expansion, recession, or recovery. Moreover, while the most common metric of these model performances is accuracy, a higher accuracy does not really translate into higher ROI, especially because the model fits well with just one type of instrument or one market condition. Therefore, most of the current methods have very high accuracy but poor financial returns and have no scope for enhancement in their mechanism.

### 1.3 Contribution

This paper is aimed at the weaknesses in several of the available algorithmic trading models, hence proposing a flexible, multi-model, multi-instrument trading system to improve accuracy and increase ROI. Unlike traditional models that focus on one single instrument or very narrow range of assets, our approach allows users to select a list of instruments and to dynamically adjust models and instruments while trading. This flexibility provides more options for trading and helps the system avoid downtrends in specific instruments or markets. The model also allows for instrument selection depending on the user's preference. This will be useful in cases where the use of certain instruments is restricted by regulations-such as for some companies or in some countries. This extends the system to more users and further extends its usability across various regulatory frameworks.

This system, by incorporating a number of models and instruments, is more capable of adapting to different market conditions and behaviors of instruments. This two-layer algorithm trains multiple models of various instruments in the first layer and, in turn, uses a VotingClassifier to determine the best instrument-model pairs in the second layer according to a range of performance metrics including accuracy, precision, and backtest result. The essence of this multi-instrument, multi-model approach is to increase flexibility for the system to adapt into different market phases and therefore to enhance predictive accuracy and financial returns.

This paper's contribution in one major way is that it will allow the users themselves to decide on instruments, embedding user reliance in it. This will not only enhance user flexibility but also widen the horizons of the system itself-that is, more options to trade and adapt to various market conditions mean overall improvement in the performance of the system. Another important aspect is the prevention of overfitting in this system, which includes a set of performance metrics beyond just accuracy: precision, recall, F1-score, AUC, and results of backtests[4] which are also used in other researches [5]. These give further view on model performance in a more profound way, allowing to ensure high accuracy does not involve poor financial returns.

It is also designed to accommodate future development. The model continuously allows for the addition of new datasets and models, which also improves overall performance over time. With more instruments and further models added in, such a system can continue to adapt and improve toward trading. Thus, it is a solid, scalable platform suitable for algorithmic trading.

Namely, this is a novel, flexible, and highly customizable trading system that will incorporate multiple models and instruments to achieve better accuracy and ROI. It is capable of handling different market conditions, as it allows the user to select instruments against suitability for their purposes or even regulatory requirements. Finally, the system architecture is open and developable on a continuous basis dynamic solution for ever-evolving market environments.

The paper is organized as follows: in the next sections, some related works will be reviewed, pointing out their advantages and limitations; then a detailed description of the algorithm proposed will be given, with its multi-pair and multi-layer architecture. Finally, the algorithm's results are going to be presented, and conclusions based on the findings and possible future directions are summarized.

# 2 RELATED WORKS

The section provides a review of the literature for several machine learning-based algorithmic trading methodologies proposed and implemented. It discusses the most relevant approaches, highlighting their similarities and comparing methodologies and outcomes for the selected papers.

## 2.1 Algorithmic trading with directional changes

Recent works have considered multi-strategy approaches, which show fair potential to improve trading performance in various financial markets. Paper [6] introduces a novel DC-based trading strategy based on a genetic algorithm for finding the best combination among multiple DC-based strategy recommendations. Therefore, it compares a multi-threshold DC strategy with five single-threshold DC strategies, three technical analysis indicators, and a buy-and-hold strategy. With this strategy, MTDC returns 1.15% per month, using 200 monthly datasets from 20 Forex markets, more than a two-fold outperformance of the best single threshold return of 0.53%. Similarly, the present paper also follows a multi-strategy approach. It makes use of a voting classifier to choose the most profitable pairs of instruments and models. Though paper [6] has used 10-minute interval data, this paper applies daily interval data. Though data has been analyzed at different intervals in this study, it reaches a return of 2.5% over a three-month period.

Moreover, Paper [6] was executed on a non-dedicated Red Hat Enterprise Linux system with 24 cores and 24 GB of memory, where the training process took 330 minutes to complete. In contrast, the execution in this paper utilized Apple Silicon ARM64 with M2 architecture, finishing training in 80 minutes. While the two studies differ in terms of the number of datasets and intervals used, the return on investment (ROI) perspective demonstrates a significant improvement in computational efficiency in this work. Despite the hardware and dataset differences, this paper offers a more efficient execution time while achieving comparable or better returns, highlighting the advancements in computational performance. While paper [6] suggests future research to dynamically choose the number of thresholds for each dataset, this paper has indeed implemented a flexible strategy selection mechanism using the voting classifier, which allows dynamic selection of both instruments and models. Thus, it is far more adaptable to many changing market conditions. This paper, just like the paper [6], underlines the advantages of multi-strategy optimization: in both cases, such models outperformed any single-strategy model both by profitability and risk management.

## 2.2 Candlestick patterns and sequence similarity

This may provide a basis of comparison, as different algorithmic trading studies use a diversity of metrics when evaluating against others. The authors in paper [7] propose the multivariate financial time series stock forecasting model that integrates sequential pattern mining and sequence similarity. This model improves the accuracy of the stock trend forecast by using K-line pattern mining on multivariate time-series data. Experimental investigations were conducted on financial time series data of the constituents of the CSI 300 and CSI 500. The proposed hybrid model achieved the average accuracy of 56.05% and 55.56% for two different datasets, while SVM and LSTM models achieved 51.83% and 51.32%, and 50.71% and 50.68%, respectively. Specifically, this project has some similarities with paper [7] in the type of data applied, evaluation metrics, and objectives. However, all critical preprocessing stages and machine learning models are very different.

From the system structure point of view, paper [7] developed a model for predicting the direction of the next candle, which is the same goal as this project. However, relying solely on accuracy may not work as effectively in algorithmic trading. Candles differ in return: some candles, like doji candles, have negligible returns, whereas a single candle often results in a return of more than 10% price change. Thus, for two instruments, the results in paper [7] were more accurate. This paper uses a composition of machine learning layers with an ensemble model for instrument-model pairs detection and improves accuracy on the prediction itself and most importantly by enhancing returns, crucial for trading strategies.

## 2.3 Other related works

While such methods are indeed adequate in terms of achieving a high degree of accuracy, they do not meet the requirements when it comes to ROI. All works in [7], [8], [9], [10] including [11] present algorithms of single instrument prediction, where one model is tuned for one instrument. Contrasting that, the paper provides a better solution that allows multiple instruments to be handled; thus, treating them as a portfolio. In addition, the suggested model is flexible, because it is easy to include new models for example [7], [8], [9], [10] and [11] strategies in the system.

## 3 Approach

This section introduces the approaches used for the inputs, outputs, and core functions of this project. Every algorithmic trading system contains the same building blocks: data fetching, preprocessing, core model development, backtesting, and results presentation[12]. In this project, the two-layer multi-model machine learning architecture is emphasized to ensure generality. The flexibility of the trading system allows for broad applicability across financial instruments and enhances traditional investment methodologies. This also provides a base for other projects, functioning like an internal search engine to find the most suitable instruments and ideal models.

### 3.1 Data representation and description

The time series data for this project is sourced from the yfinance library and OANDA broker APIs, encompassing price data for supported symbols. Each dataset, as shown in Table 1, comprises six quantitative features: Open, High, Low, Close, Adj Close, and Volume [13].

Table 1. Input Data

| Date | Open | High | Low | Close | Adj Close | Volume |
| --- | --- | --- | --- | --- | --- | --- |
| 2013-12-24 | 1199.800049 | 1205.599976 | 1197.699951 | 1205.099976 | 1205.099976 | 184 |
| 2013-12-26 | 1207.099976 | 1215.900024 | 1207.099976 | 1214.099976 | 1214.099976 | 140 |
| 2013-12-27 | 1213.400024 | 1218.500000 | 121.9000024 | 1216.099976 | 1216.099976 | 278 |
| 2013-12-30 | 1215.00000 | 1215.000000 | 1194.400024 | 1203.099976 | 1203.099976 | 351 |
| … | … | … | … | … | … | … |

During each training cycle, a list of financial instruments is fetched from the data provider. Based on the method used from the instrument-model search part of the algorithm, generally, more instruments will yield better results concerning the ROI [1]. On the other hand, this also requires increased computational power, which consequently slows down the training phase.

### 3.2 Feature Extraction and Pre-processing

In the preprocessing step, new features are generated for each instrument, including return, label, and indicators. Table 2 demonstrates all added features and their corresponding calculation formulas.

An indicator refers to a statistical calculation or measurement used to assess and analyze various aspects of financial markets, assets, or economic conditions. Indicators are often derived from financial data and are employed by traders, analysts, and investors to gain insights into market trends, potential price movements, and overall economic health. These indicators can cover a wide range of metrics, including price levels, trading volumes, volatility, and other relevant factors. In this project, the following indicators are utilized:

Moving averages: Moving averages are commonly used to smooth out price data and identify trends over a specific period. The Simple Moving Average (SMA) is calculated by summing up a set of prices over a specified period and dividing by the number of data points [14].

Relative Strength Index: RSI is a momentum oscillator that measures the speed and change of price movements. RSI values range from 0 to 100 and are often used to identify overbought or oversold conditions [15].

Moving Average Convergence Divergence: MACD is a trend-following momentum indicator that shows the relationship between two moving averages of an asset's price. It consists of the MACD line, Signal line, and Histogram [16].

After adding additional features, each data point will be shifted based on its previous point, creating a time series data structure similar to the one generated by the TimeseriesGenerator function, which will be employed for Keras preprocessing.

Labels are categorical values representing the project output, taking on values of 0 and 1. A label is assigned the value of 1 if the return on the given day is positive and 0 otherwise.

Table 2. Features Formulas

| Features | Formula |
| --- | --- |
| Return | $(Close - Open) / Open \times 100$ |
| SMA | $Number\ of\ Periods / Sum\ of\ Prices$ |
| RSI | $100 - (100 / 1+RS)$ |
| MACD Line | $(12-dayEMA) - (26-dayEMA)$ |
| Signal Line | $9-dayEMA\ of\ MACDLine$ |
| Histogram | $MACDLine - SignalLine$ |

## 3.3 Partitioning steps

Following the data collection and preprocessing steps, the dataset will be divided into model development and evaluation. Since this is a short-term prediction algorithm, the data will be split into a training set and test set using a 95-5% split, with 95% going to training. The remaining 5% would be kept exclusively for the final evaluation of the model, while the training set would further be divided into learning and validation sets in a 90-10 ratio.

## 3.4 Composing Ensembles of Instruments-Models

The system structure consists of a learning phase, validation evaluation, and the detection of successful models among all trained models for all instruments. Let's formalize this as a function composition, $Prediction = (V\ o\ E\ o\ M)(x)$ where $M$ is the first learning layer, $E$ is the evaluation of $M$, and $V$ is the second layer, which selects the best instrument-model pairs based on the evaluation metrics of the models. Here, $x$ represents the input: a list of instruments along with their associated time series data. The final prediction will be the selected instrument and its pre-trained model. The input to the VotingClassifier will be a 3-dimensional matrix, this tensor will have dimensions $i * j * k$, where $i$ represents the number of instruments, $j$ represents the number of models, and $k$ corresponds to the evaluation metrics.

$$Prediction = VotingClassifier\left(\begin{bmatrix}E_1\\E_2\\...\\E_k\end{bmatrix}\left([M_1\quad M_2\quad ...\quad M_j] * \begin{bmatrix}x_1\\x_2\\...\\x_i\end{bmatrix}\right)\right)$$

### 3.4.1 Training Classification Models with Instruments

In the learning phase, each of the classification models listed in Table 3 will be trained with a list of datasets belonging to different instruments. The result will be a number of instrument-model pairs equal to the product of the number of models and the number of instruments. The training models are hyperparameter-tuned using the GridSearch method to enhance their overall accuracy.

Table 3. List of Machine-Learing Models

| |
|---|
| GradientBoostingClassifier(n_estimators=100, learning_rate=0.01, max_depth=12)} |
| LogisticRegression(C=0.1, max_iter=10000, solver='lbfgs')} |
| DecisionTreeClassifier(max_depth = 12, min_samples_split=6, min_samples_leaf=4)} |
| RandomForestClassifier(n_estimators = 500, max_depth = 10)} |
| KNeighborsClassifier(n_neighbors = 7, weights='distance')} |
| GaussianNB(var_smoothing=1e-09)} |
| LinearSVC(C=1 ,max_iter = 10000)} |
| MLPClassifier(hidden_layer_sizes = (69),max_iter = 10000 ,activation='relu', alpha=0.0001)} |
| SupportVectorClassifier(C=1, kernel='rbf', gamma='scale', max_iter=5000)} |

### 3.4.2 Evaluating Pairs

Following the learning phase, the models undergo validation with dedicated datasets, where their performance is assessed using various metrics such as accuracy, backtest, nnp, NormalizedAcc, precision, recall, f1, and auc. These metrics play a crucial role in evaluating the effectiveness of the models and guiding the training of new machine-learning model to identify a successful classification model and its instrument to be utilized.

All metrics are sourced from sklearn.metrics, except for backtest, nnp, and NormalizedAcc. The backtest metric represents the output of a function that displays the return of the validation dataset using the considered model. Nnp signifies the normal return profit without any machine learning assistance. NormalizedAcc represents normalized accuracy calculated using the following formula:

$$NormalizeAcc = counts[1]/(counts[0]+counts[1])$$

Certainly, considering the potential issue with models setting all outputs to 0 or 1 to achieve higher accuracy in the presence of imbalanced labels, the formula for NormalizedAcc can be adapted to address this challenge. NormalizedAcc provides a more balanced evaluation, considering the baseline accuracy that could be achieved by random guessing.

As a result of the initial phase of training, a data frame is generated that illustrates the effectiveness of models for each symbol. In this study, the results are shown for 13 symbols and 11 models, producing a total of 141 rows.

- $M_j$ represent the i-th model where i = 1,2,…,11.
- $D_i$ represent the j-th dataset where j = 1,2,…,13.
- $E_{ij}$ represent the evaluation metrics for model $M_j$ on dataset $D_i$.
- $k$ represents the number of evaluation metrics (e.g., accuracy, precision, recall, F1-score).

We can define the evaluation for each model-dataset pair as a matrix $E$, where each $E_{ij}$ is a vetor of size $k$ (evaluation metrics for the j-th model on the i-th dataset).

$$E = \begin{bmatrix} E_{11} & E_{12} & \cdots & E_{1j} \\ E_{21} & E_{22} & \ddots & \vdots \\ \vdots & \vdots & \vdots & \vdots \\ E_{i1} & E_{i2} & \cdots & E_{ij} \end{bmatrix}$$

Where each $E_{ij} = [e_{ij}^1, e_{ij}^2, \ldots, e_{ij}^k]$, representing the set of evaluation metrics for model $M_j$ on dataset $D_i$.

The evaluations are stored in a Data Frame where the rows correspond to the (model, dataset) pairs and the columns correspond to the evaluation metrics.

$$Second\ Layer\ Input = \begin{bmatrix} Dataset & Model & Metric_1 & Metric_2 & \cdots & Metric_k \\ D_1 & M_1 & e_{11}^1 & e_{11}^2 & \cdots & e_{11}^k \\ D_1 & M_2 & e_{12}^1 & e_{12}^2 & \cdots & e_{12}^k \\ \vdots & \vdots & \vdots & \vdots & \ddots & \vdots \\ D_i & M_j & e_{ij}^1 & e_{ij}^2 & \cdots & e_{ij}^k \end{bmatrix}$$

An example of the first layer output and second layer input can be visualized in Table 4, which, once partitioned, is ready to be fed into the second layer model, a VotingClassifier. This model identifies profitable instruments along with their pre-trained machine learning models.

Tabel 4. Input of Second-Layer training

| Dataset | Model | Accuracy | NormalizeAcc | Precision | … | Profit % | Label |
|---|---|---|---|---|---|---|---|
| AAPL | LogisticRegression | 0.519078 | 0.492486 | 0.560150 | … | -0.7686 | 0 |
| AAPL | DecisionTreeClassifier | 0.504680 | 0.478824 | 0/553140 | … | -0.8717 | 0 |
| … | … | … | … | … | … | … | … |
| GC=F | SVC | 0.532084 | 0.504098 | 0.520000 | … | -0.5121 | 0 |
| GC=F | MLPClassifier | 0.488825 | 0.463115 | 0.471992 | … | -0.4681 | 0 |

### 3.4.3 Detecting Profitable Pairs with Voting Classifier

Since trading strategies must be able to generate sufficient returns to cover their associated costs with some amount of certainty if they are to warrant an allocation[1], this layer performs the identification of financial instruments with high profit potential and their respective models based on the output obtained from a VotingClassifier model trained on the historical performance data provided from the previous layer, which has persisted in a MongoDB database. In each further training phase, the second layer is retrained on more and more data, each time with higher accuracy. Therefore, the algorithm improves with each increment of data provided in the continuous learning and optimization process.

What is the real contribution of the second layer? So far we have models, some of them are profitable, others not. The accuracy of the second layer is crucial to determining the general performance. Suppose that the accuracy of our profitable models is 0.51 and the less profitable models are 0.49. Assuming that the accuracy of the second layer is 0.80, then the general accuracy can be expressed as:

$$Mean\ Accuarcy = \sum_{i=0}^{M} V_i\ P(V_i) = +1(80\%) - 1(20\%) = 60\%$$

Therefore, the second layer with its 0.80 accuracy generally improves the accuracy of the system to 0.60.

In the implementation part, which will be discussed further, the output, which consists of potential successful models, will be implemented on the OANDA API. These models will predict the direction of their linked instruments for the following day.

### 3.5 Implementation Details

This section encompasses the implementation of the project with the OANDA API, consolidated into a single cell. The script is designed to be implemented in a cloud environment, running continuously 24/7. It incorporates a sleep function after code execution, allowing it to rest for a day. Additionally, the script operates exclusively between 22:00 and 23:59 hours.

## 4 Evaluations and Results

To obtain the project's results, first, the Table 5 instruments are employed:

Table 1. List of assets used as inputs

| Ticker Symbol | Financial Instruments, Assets |
|---|---|
| AAPL | Apple Inc. |
| GOOGL | Alphabet Inc. (Google) |
| AMZN | Amazon.com Inc. |
| TSLA | Tesla, Inc. |
| META | Meta Platforms, Inc. (formerly Facebook) |
| CSCO | Cisco Systems, Inc. |
| NVDA | NVIDIA Corporation |
| NFLX | Netflix Inc. |
| JPM | JPM - JPMorgan Chase & Co. |
| IBM | International Business Machines Corporation |
| BTC-USD | Bitcoin to US Dollar |
| ETH-USD | Ethereum to US Dollar |
| GC=F | Gold |

Among all the selected symbol-model pairs, those chosen by the best-model-finder are going to be tested on the test dataset. The best-model-finder employs two selection methods: one for choosing the best symbol-model, and the other for selecting lists of symbol-model pairs predicted to be profitable. This method is set manually. In Figure 2, four subcharts exist, each trained at different times. This adjustment is made by selecting the best symbol-model for each specific time period. The "StrategyReturns," depicted by the green line, represents the return of selected symbols enhanced by their chosen machine learning models. On the other hand, the "NormalReturns," illustrated by the blue line, signifies the return of selected symbols without employing any machine learning techniques.

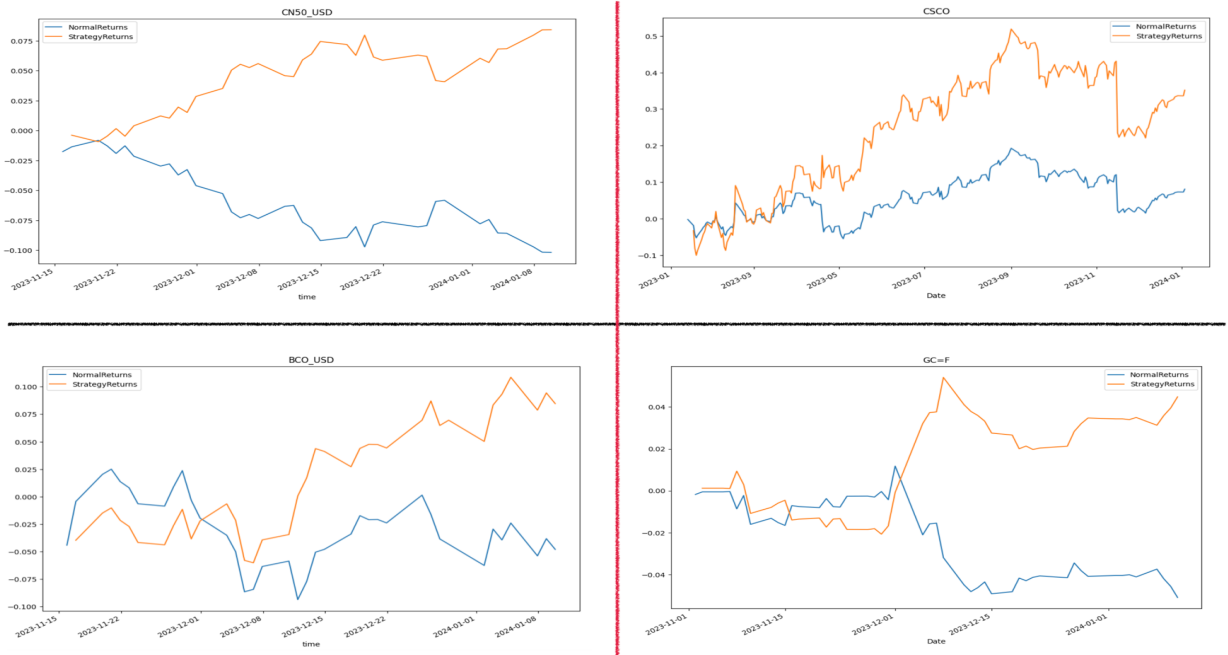

Figure1. Results of single instrument (best pair)

In the Figure 3, there are four subcharts, each trained at different times with various instruments. This adjustment is made by selecting lists of symbol-models for each specific time period. The "ModelsReturn," depicted by the blue line, represents the return of selected symbols enhanced by their chosen machine learning models. On the other hand, the "NormalReturn," illustrated by the green line, signifies the return of selected symbols without employing any machine learning techniques.

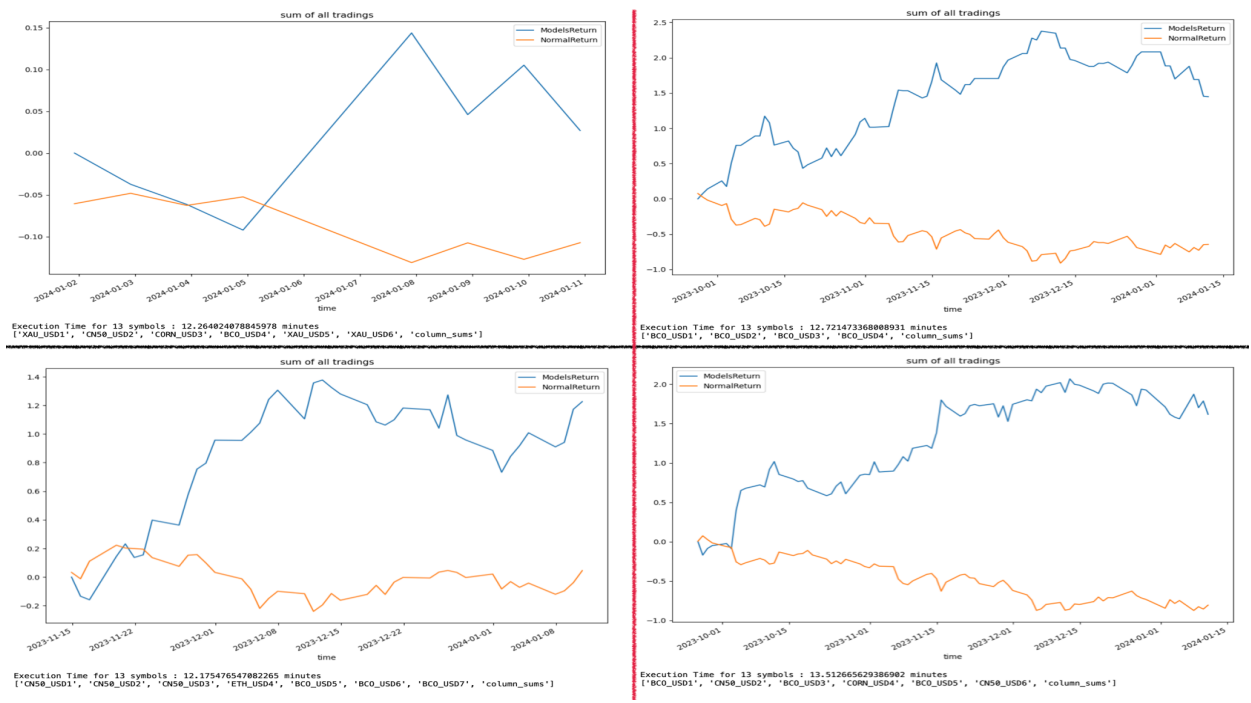

Figure 2. Result of choosen instrument-model pairs

Figure 4 and Figure 5 are selected to show the ROI of the method presented in paper [8], implemented over both long-term and short-term periods. By this, it may be concluded that although the mean accuracy of the approach in this paper is lower than that of paper [11], yet the ROI is much higher. The best part is that all this can be done in a very short period, say one month, as it allows great flexibility regarding the choice of instruments and strategies.

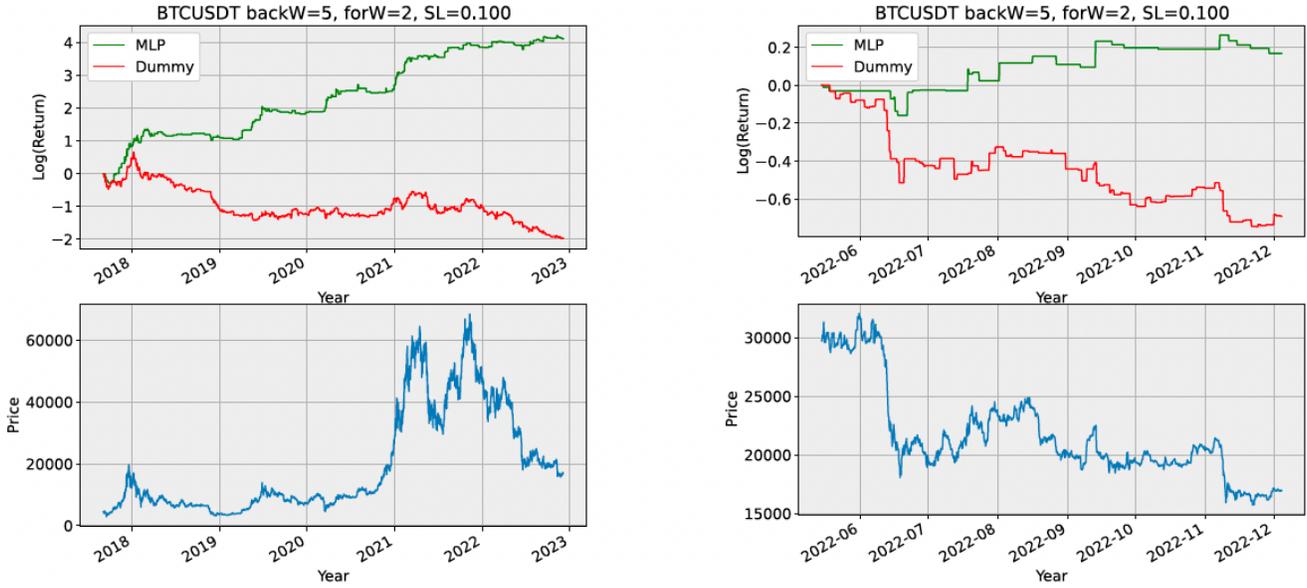

Figure 3. BTC-USDT Logarithmic return [11]

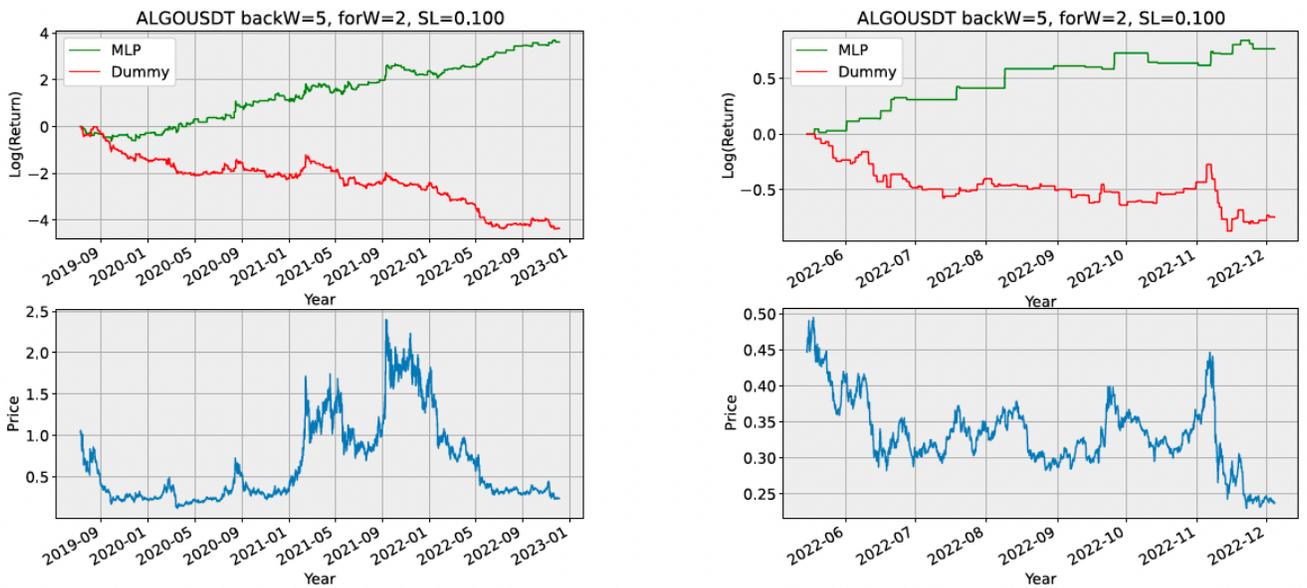

Figure 4. ALGO-USDT Logarithmic return [11]

# 5 CONCLUSION

Results obtained in the experiment testify to high efficiency of the approach being proposed. Currently, the project has already been tested on real-world data provided by OANDA broker, and computations are performed on the Apple Silicon ARM64 with M2 architecture. This underlines the possibility of obtaining promising results even with managing computational resources. Besides, this project has great potential for possible future improvements. Since the system works like a search engine, constantly looking for an optimal opportunity, it relies heavily on having a flow of data at all times. Further development of the project could be done by increasing the number of financial instruments included, expanding the list of models like CNNs, RNNs and NLP models[17], [18], or implementing other preprocessing methods into the algorithm, such as advanced oscillators. Improvements might also be made by incorporating more varied and enriched data sources, including financial news, oscillators, reliable model outputs, or correlations between these factors. Though the data and resources are at a premium, the models could still find promising opportunities to lay a foundation for further work.